\documentclass[12pt]{article}

\usepackage{enumitem}
\usepackage[tmargin=2cm,lmargin=2.5cm,bmargin=3cm,hmarginratio=1:1,headheight=65pt,headsep=1cm]{geometry} 
\usepackage[T1]{fontenc}
\usepackage[french,english]{babel}
\usepackage{color,graphicx}
\usepackage{amsmath,amssymb,amsfonts}
\usepackage{amsthm}
\usepackage{mathtools}
\usepackage{slashed}
\usepackage{sidecap} 
\usepackage{array}
\usepackage[indentafter]{titlesec}
\usepackage[utf8]{inputenc}
\usepackage{setspace}
\usepackage{perpage} 
\usepackage{fancyhdr} 
\usepackage{emptypage}
\usepackage{pifont} 
\usepackage{cite} 

\usepackage{hyperref} 

\usepackage{color}

\newcommand{\badat}{\begin{alignedat}}
\newcommand{\eadat}{\end{alignedat}}

\date{}
\begin{document}

\begin{titlepage}
  \thispagestyle{empty}
  \begin{center}  
\null 
\null

{\LARGE\textbf{A note on the spectral flow operator}}

\vskip1.2cm
   \centerline{Gaston Giribet}
\vskip1cm

{Abdus Salam International Centre for Theoretical Physics, ICTP}\\
{{\it Strada Costiera 11, Trieste, 34151, Italy.}}

{Physics Department, University of Buenos Aires and IFIBA-CONICET}\\
{{\it Ciudad Universitaria, Pabell\'on 1, Buenos Aires, 1428, Argentina.}}

\end{center}

\vskip1cm

\begin{abstract}
The insertion of the spectral flow operator in a string scattering amplitude on AdS$_3\times \mathcal{N}$ produces a change in the winding number of one of the incoming (or outgoing) states, making it possible to compute amplitudes of processes in which winding number in AdS$_3$ is not conserved. The insertion of such operator, however, might seem artificial from the worldsheet theory perspective, as it appears as an unintegrated vertex operator of conformal dimension zero that does not represent any normalizable state. Here, we show that the spectral flow operator naturally emerges in the Liouville field theory description of the WZW correlation functions once it is combined with a series of duality relations among conformal integrals. By considering multiple insertions of spectral flow operators, we study the dependence on the moduli for arbitrary number of them, and we show explicitly that the amplitude does not depend on the specific locations of the accessory insertions in the worldsheet, as required by consistency. This generalizes previous computations in which particular cases were considered. This can also be thought of as an alternative proof of the WZW-Liouville correspondence in the case of maximally winding violating correlators.
\end{abstract}

\end{titlepage}

\section{Introduction}

String theory on AdS$_3\times \mathcal{N}$ with pure NS-NS fluxes provides an excellent arena to test AdS/CFT correspondence beyond the supergravity approximation \cite{GKS, GS, Ooguri}. The worldhseet theory on AdS$_3$, being described by the $SL(2,\mathbb{R})_k$ Wess-Zumino-Witten (WZW) model, can be quantized and in principle solved exactly. This permits to have access to the finite $k=R^2 /\alpha ' $ regime of the theory, in which the size of the strings, $\ell_s=\sqrt{\alpha '}$, is not necessarily large in comparison to the radius of the space, $R$. This enables to find the spectrum of the quantum theory in terms of $\hat{sl}(2)_k$ Kac-Moody unitary representations \cite{MO1,GN2} and to compute scattering amplitudes by integrating the WZW correlation functions \cite{GN1, GN3, MO3}.

The results for string amplitudes in AdS$_3$ \cite{BB, Teschner1, Teschner2}, once combined with non-renormalization theorems that are now available \cite{deBoer:2008ss}, permitted in \cite{Gaberdiel:2007vu, Dabholkar:2007ey, Pakman:2007hn, Giribet:2007wp, Cardona:2009hk} to perform precision checks of AdS/CFT correspondence at large $N$ with $k$ finite. The tree level 3-point functions of protected chiral operators in type IIB superstring theory on AdS$_3\times S^3 \times T^4$ with NS-NS flux were computed in \cite{Gaberdiel:2007vu, Dabholkar:2007ey} and were shown to exactly reproduce the computations in the symmetric product CFT$_2$ at the orbifold point in the large $N$ limit. The analysis was then extended to the chiral $\mathcal{N}=4$ operators in \cite{Pakman:2007hn}, and it was later completed in \cite{Giribet:2007wp, Cardona:2009hk} by adding the winding string sectors, which correspond to spectrally flowed representations of $\hat{sl}(2)_k$.

In the last two years, the interest on strings on AdS$_3$ has been renewed. Special attention has recently been focused on the superstring theory at the point $k=1$, where special features appear \cite{Giveon:2005mi, Giribet:2004zd, Gaberdiel:2017oqg, Ferreira:2017pgt}. Different proposals for the holographic description of the supersymmetric theory at $k=1$ have recently been proposed \cite{Giribet:2018ada, Gaberdiel:2018rqv, Eberhardt:2018ouy, Eberhardt:2019qcl, Eberhardt:2019niq}. In all such proposals the winding string sectors play a fundamental role. Here, we will study the theory for arbitrary $k$; we will focus on the bosonic theory and analyze the winding string sectors in detail. Our main goal is to clarify the prescription for computing winding violating string scattering amplitudes proposed in \cite{MO3} and make it easy to understand from the worldsheet CFT$_2$ point of view.

The paper is organized as follows: In section 2, we concisely review string theory on AdS$_3$ with NS-NS fluxes. We study the spectrum and the definition of tree level scattering amplitudes. Since we will be mainly involved with winding violating amplitudes, we will need to introduce the so-called spectral flow operator, which will be the principal ingredient in our discussion. In section 3, we present the most salient properties of the spectral flow operator in relation to correlation functions, and we study the 3-point function in detail. In section 4, the main tool to solve the amplitudes is presented: This is the so-called $H_3^+$-Liouville correspondence, which is a dictionary that maps $SL(2,\mathbb{R})_k$ WZW correlation functions into correlation functions in Liouville field theory (LFT). In section 6, we analyze the integral representation of correlation functions in LFT together with a series of duality relations among different conformal integrals that will ultimately lead to compute the relevant 3-point functions. In section 7, we comment on the $n$-point functions.

\section{String amplitudes on AdS$_3$}

We will be concerned with tree-level string amplitudes on AdS$_3 \times \mathcal{N}$ with pure NS-NS fluxes. These observables are given by
\begin{equation}
\mathcal{A}^n_{{\bf p}_1,{\bf p}_2,...{\bf p}_n} = \int {\prod_{i=1}^{n} d^2z_i}\   \frac{{C}_{\text{AdS}_3}^{n}(z_1 , ... , z_n ) \times {C}_{\mathcal{N}}^{n}(z_1 , ... , z_n )}{\text{Vol}(PSL(2,\mathbb{C})) } \label{AB}
\end{equation}
where the integrals are over ${\mathbb{CP}^1}$; ${C}_{\text{AdS}_3}^{n}$ is the $n$-point correlation function of primary operators in the $SL(2,\mathbb{R})_k$ WZW model on the Riemann sphere, 
\begin{equation}
{C}_{\text{AdS}_3}^{n}(z_1 , ... , z_n )=\Big\langle  \prod_{i=1}^n \Phi_{j_i,m_i,\bar{m}_i}^{\omega_i}(z_i) \Big\rangle_{sl(2)}^{\sum_{i=1}^n\omega_i} , \label{inn}
\end{equation}
which describes the string $\sigma $-model on AdS$_3$; ${C}_{\mathcal{N}}^{n}$ is the contribution of the internal CFT on $\mathcal{N}$. In (\ref{inn}), we are omitting normal ordering symbols. Subscript $sl(2)$ on the right hand side makes explicit that (\ref{inn}) is a correlator in the non-compact $SL(2,\mathbb{R})$ model. Superscript $\sum_{i=1}^{n}\omega_i$ indicates the total winding number in the correlator, as in AdS$_3$ such quantity is not necessarily conserved \cite{GN3, MO3}. The indices ${\bf p}_i$ in (\ref{AB}) represent the momenta of the incoming and outgoing states. Both ${C}_{\mathcal{N}}^{n}$ and ${C}_{\text{AdS}_3}^{n}$ depend on ${\bf p}_i$, although we do not write explicitly such dependence for short. The volume of the conformal Killing group, $\text{Vol}(PSL(2,\mathbb{C}))$, can be canceled by fixing on the punctured projective complex plane three of the $n$ points at which the vertices are inserted; as usual, we choose $z_1=1-z_2=1/z_3=0$.

The spectrum of the theory on AdS$_3\times \mathcal{N}$ was worked out by Maldacena and Ooguri in \cite{MO1}. Let us briefly review it here: Operators $\Phi_{j,m,\bar{m}}$ in (\ref{inn}) represent the AdS$_3$ part of the string vertex operators that create Virasoro primary states in the worldsheet CFT. These states organize themselves in representations of the $\hat{sl}(2)_k\oplus \hat{sl}(2)_k$ affine Kac-Moody algebra. These representations are built out of unitary, Hermitian representations of $SL(2,\mathbb{R})$. More precisely, the Hilbert space of the theory is constructed by acting with elements of the enveloping algebra of $\hat{sl}(2)_k$ on states of suitable representations of $SL(2,\mathbb{R})$. Such representations are the highest- and lowest-weight discrete series $\mathcal{D}^{\pm }_{j}$, and the principal continuous series $\mathcal{C}_j^\lambda $ of $SL(2,\mathbb{R})$, together with their spectrally flowed images $\mathcal{D}^{\pm ,\omega }_{j}$ and $\mathcal{C}^{\lambda ,\omega }_{j}$; see \cite{MO1} for details. 

The WZW level is given by $k=R^2/\alpha '$; so it controls the size of the strings relative to the size of AdS$_3$ space, implying that the semiclassical limit corresponds to the limit $k\to \infty $. The central charge of the worldsheet CFT takes the form
\begin{equation}
c=\frac{3k}{k-2}+c_{\mathcal{N}}\equiv 26,
\end{equation}
where $c_{\mathcal{N}}$ is the contribution of the internal CFT on $\mathcal{N}$. This sets the level to be $k=2(c_{\mathcal{N}}-26)/(c_{\mathcal{N}}-23)$; as expected, it yields $c_{\mathcal{N}}-26=3$ in the large $k$ limit, in which the 3-dimensional AdS space is softly curved.

As said, the indices ${\bf p}_i$ represent the labels that parameterize the momenta of both the AdS$_3$ and the internal parts; in the AdS$_3$ part these labels are those that classify the $SL(2,\mathbb{R})\times SL(2,\mathbb{R})$ representations: $j_i$, $m_i$, $\bar{m}_i$, and $\omega_i$. {Discrete representations of the universal covering of $SL(2,\mathbb{R})$, $\mathcal{D}^{\pm  }_{j}$, correspond to $j\in \mathbb{R}$, $m\in \pm j\pm \mathbb{Z}_{\geq 0}$, and $\omega =0 $; while continuous representations, $\mathcal{C}^{\lambda }_{j}$, correspond to $j\in 1/2+i\mathbb{R}$, $m\in \lambda +\mathbb{Z}$, and $\omega =0$. The sectors $\omega \neq 0$ are defined as the Kac-Moody primaries with respect to the algebra generators obtained by transforming the original ones with a $\mathbb{Z}$-valued spectral flow isomorphism \cite{MO1}, with $\omega \in \mathbb{Z}$ being the spectral flow parameter. The set of unitary, Hermitian representations of $SL(2,\mathbb{R})$ also includes the complementary series $\mathcal{E}_j^\alpha$, but these series are not necessary to construct the string spectrum\footnote{There are two arguments to exclude the complementary series $\mathcal{E}_j^\alpha$ from the string spectrum. On the one hand, the 1-loop partition function of the theory defined only with the series $\mathcal{C}^{\lambda ,\omega }_{j}$ and $\mathcal{D}^{\pm  ,\omega }_{j}$ results to be modular invariant \cite{MO2}. On the other hand, the states of the series $\mathcal{C}^{\lambda ,\omega }_{j}$, $\mathcal{D}^{\pm  ,\omega }_{j}$ form a basis of the $L^2$ functions \cite{MO1}.}. The spectrum also excludes discrete representations above certain value $j_{\text{max}}=(k-1)/2 $ in virtue of the strong version of the no-ghost bound \cite{MO1}.} More precisely, one only considers discrete representations in the segment 
\begin{equation}
\frac{1}{2} \leq j \leq \frac{k-1}{2}. \label{Unita}
\end{equation}

The physical interpretation of the $SL(2,\mathbb{R})$ labels $j$, $m$, $\bar m $, and $\omega$ is the following: The energy of the string states in AdS$_3$ is given by the combination $E\equiv m+\bar m + k\omega \in \mathbb{R}$, where $\omega \in \mathbb{Z}$ is the winding number --recall that $k$ is proportional to the string tension--. The quantity $J\equiv m-\bar m\in \mathbb{Z}$ gives the angular momentum in AdS$_3$, and the imaginary part of the variable $j$ can be associated to the radial momentum. 

The worldsheet conformal dimension of the primary operators on AdS$_3\times \mathcal{N}$ read\footnote{These formulas are invariant under $j\to 1-j$ and $m,\bar m , \omega \to -m,-\bar m , -\omega $. The latter can be regarded as a CPT transformation; states with negative values of $m$, $\bar{m}$, and $\omega$ have to be regarded as outgoing states.}
\begin{equation}
h=\frac{j(1-j)}{k-2}-m\omega -\frac{k}{4} \omega^2+h_{\mathcal{N}} + N  , \  \ \ h=\frac{j(1-j)}{k-2}-\bar{m}\omega -\frac{k}{4} \omega^2+\bar{h}_{\mathcal{N}} + \bar{N} \label{L3}
\end{equation}
where $h_{\mathcal{N}}$, $\bar{h}_{\mathcal{N}}$ represent the conformal dimensions of the $\sigma $-model on $\mathcal{N}$, and where $N$, $\bar{N}$ indicate the oscillator numbers. The level matching condition thus implies ${h}_{\mathcal{N}}-\bar{h}_{\mathcal{N}}\in \mathbb{Z}$. By imposing the Virasoro constrains on (\ref{L3}), we obtain the energy spectrum of the theory,
\begin{equation}
E=\frac{k }{2} \omega +\frac{1}{\omega}\Big( 2\frac{j(1-j)}{k-2} + N_{\mathcal{N}}+ \bar{N}_{\mathcal{N}}-2+h_{\mathcal{N}}+\bar{h}_{\mathcal{N}}\Big).
\end{equation}
Spectrally flowed discrete representations $\mathcal{D}^{\pm ,\omega }_{j\in \mathbb{R}}$, which describe short strings confined in the bulk of the space, have discrete energy spectrum. In contrast, continuous representations $\mathcal{C}^{\lambda ,\omega }_{j\in 1/2+i\mathbb{R}}$, which describe long strings that can reach the boundary asymptotically with winding number $\omega $, have continuous energy spectrum.

It is worth mentioning that the winding number in AdS$_3$ is not a topological degree of freedom: AdS$_3$ space is simply connected. Number $\omega $ is rather associated to the presence of the NS-NS $B$-field in the bulk, to which the strings couple. Not being topological, the winding number can in principle change when interactions take place --although it is preserved in a 2-point function--. Here, we will be concerned with amplitudes of $n$-string scattering processes in which the total winding number is not conserved. This means that we will compute $SL(2,\mathbb{R})$ WZW correlation functions that include the so-called spectral flow operator \cite{GN3, MO3}; namely
\begin{equation}
{C}^{n+1}_{\text{AdS}_3}(z_1, ... , z_{n+1})=\frac{1}{Z_{\mp }}\Big\langle   \prod_{i=1}^n \Phi_{j_i,m_i,\bar{m}_i}^{\omega_i}(z_i)\ \Phi_{\frac k2,\pm \frac k2,\pm \frac k2}^{\mp 1}(z_{n+1}) \Big\rangle_{sl(2)}^0  . \label{A}
\end{equation}
These correlation functions include one extra operator with fixed momenta $j_{n+1}=k/2$, $m_{n+1}=\bar m_{n+1} = \pm k/2$, $\omega_{n+1} = \mp 1$. The superscript $0$ on the right hand side means that $\sum_{i=1}^{n+1}\omega_i={\sum_{i=1}^n\omega_i\mp 1}$ gives zero. The $n+1^{\text{th}}$ operator in (\ref{A}) does not represent an external state but is an auxiliary operator. This means that the correct interpretation of a correlator like (\ref{A}) is that of an $n$-point function in which the total winding number is violated in one unit, i.e. 
\begin{equation}
\omega \equiv \sum_{i=1}^n\omega_i = \pm 1. 
\end{equation}
This prescription to compute winding non-preserving correlators was proposed originally by Fateev, Zamolodchikov, and Zamolodchikov (FZZ) \cite{FZZ}. The idea is that, once integrated in $z_1, ...,  z_n$, correlator (\ref{A}) gives the $n$-point scattering amplitudes $\mathcal{A}^n_{{\bf p}_1,...{\bf p}_n}$ of processes that violate the conservation of the total winding number, $\omega $, in one unit. This charge condition is induced by the presence of the $n+1^{\text{th}}$ operator $\Phi_{k/2,\pm k/2,\pm k/2}^{\mp 1}$ inserted at $z_{n+1}$. The FZZ prescription is such that, unlike the other worldsheet insertions, $z_{n+1}$ is left unintegrated. This is consistent with the fact that the extra $n+1^{\text{th}}$ operator has conformal dimension zero.

It is worth emphasizing that operator $\Phi^{\pm 1}_{k/2,\mp k/2, \mp k/2}$ does not represent any normalizable\footnote{One of the properties of the state $j=k/2=\mp m= \mp \bar{m}$, created by $\Phi^{\pm 1}_{k/2,\mp k/2, \mp k/2}$, which can be useful to compute correlation functions, is that it contains a null descendant $\lim _{z\to 0}J^{\pm}_{-1}\Phi^{\pm 1}_{k/2,\mp k/2, \mp k/2}(z)\vert 0\rangle =0$.} state of the theory. In fact, the value $j_{n+1}=k/2$ violates the unitarity upper bound $j_{\text{max}}=(k-1)/2$ of the physical spectrum (\ref{Unita}). Operator $\Phi_{k/2,\pm k/2,\pm k/2}^{\mp 1}$ is rather an auxiliary tool that is introduced just to alter the total winding number in a given correlator. Such operator was originally regarded \cite{FZZ} as a ``conjugated representations of the identity'' operator $\Phi_{0,0,0}^{0}=1$. The relation with the identity operator can be understood algebraically in terms of the identity between states of the representation $\mathcal{D}^{\mp , 0}_{j}$ and states of the representation $\mathcal{D}^{\pm , \mp1}_{k/2-j}$. This is related to the spectral flow isomorphism of $\hat{sl}(2)_k$, which for the spectral flow parameter $\omega =\pm 1$ does not necessarily produce new representations\footnote{In particular, operators $\Phi^{0}_{0,0,0}$, $\Phi^{-1}_{k/2,k/2,k/2}$, and $\Phi^{1}_{k/2,-k/2,-k/2}$ share both the Cartan energy $m+k \omega /2=0$ and the quadratic Casimir $(k-2)h=0$.} but two different ways of representing the same states. This is the reason why, in \cite{GN3}, operator $\Phi^{\pm 1}_{k/2,\mp k/2, \mp k/2}$ was alternatively called ``spectral flow operator''.

\section{Spectral flow operator}

Let us now discuss the spectral flow operator in a more systematic way: In the WZW theory, we will consider the special dimension-zero local operators
\begin{equation}
\frac{1}{Z_+}\Phi _{\frac{k}{2},-\frac{k}{2}, -\frac{k}{2}}^{+1} \ , \ \ \ \ \ \frac{1}{Z_-}\Phi_{\frac{k}{2},+\frac{k}{2}, +\frac{k}{2}}^{-1}. \label{L7}
\end{equation}
where the prefactor $Z^{-1}_{\pm}$ stands for a normalization (possibly divergent) yet to be fixed. This prefactor is somehow related to the $V_{conf}$ prefactor appearing in \cite{MO3}. 

Being dimension-zero operators with non-vanishing $\omega $, the insertion of (\ref{L7}) in a correlation function suffices to change the total winding number of a given amplitude without spoiling its conformal properties. Conformal invariance demands the dimension-zero operator not to be integrated. Then, an immediate question that arises is what to do with the inserting points at which they are inserted. This is a question about the structure of the moduli space, and it is one of the questions we want to discuss in detail. In this regard, we will prove that: 
\begin{enumerate}[label=(\alph*)]
	\item The final answer for the amplitude does not depend on those insertion points, even when an arbitrary number of spectral flow operators are inserted (this is far from obvious when glancing at (\ref{A}), for instance).
\item The final answer for the amplitude only depends on the difference between the number of the operators $\Phi_{k/2,-k/2, -k/2}^{+ 1}$ and the number of the operators $\Phi_{k/2,+ k/2,+ k/2}^{ -1}$ present in the correlator, regardless the net number of them.
\end{enumerate}

Showing (a) and (b) requires working out the expressions for the $SL(2,\mathbb{R})_k$ WZW correlation functions explicitly, and this is what we will do in the next sections.


The 3-point functions are the building block of the higher correlation functions. Therefore, we will focus on the following class of objects
\begin{eqnarray}
\tilde{C}^{\ 3,  n_+ , n_-}_{\text{AdS}_3}=\frac{1}{Z_+^{n_+}Z_-^{n_-}}\Big\langle \prod_{i=1}^{3} \Phi_{j_i,m_i,\bar{m}_i}^{\omega_i}(z_i) \prod_{a=1}^{n_+} \Phi_{\frac k2,-\frac k2,-\frac k2}^{+1}(u^+_a) \prod_{A=1}^{n_-} \Phi_{\frac k2,+\frac k2,+\frac k2}^{-1}(u^-_A) \Big\rangle_{sl(2)}^{0}\label{tormentita}
\end{eqnarray} 
which include an arbitrary number ($n_++n_-$) of spectral flow operators. The tilde on $\tilde{C}^{\ 3,n_+ , n_-}_{\text{AdS}_3}$ is there to remind us of the presence of such operators. Worldsheet insertion points are fixed at $z_1=0, z_2=1, z_3=\infty$. The superscript $0$ on the right hand side of (\ref{tormentita}) indicates that the total winding number vanishes, i.e. $\sum_{i=1}^{3+n_+-n_-}\omega_i=\sum_{i=1}^{3}\omega_i +n_+ - n_-=0$. As we discussed, according to the FZZ prescription, correlator (\ref{tormentita}) actually represents a winding violating $3$-point amplitude in which 
\begin{equation}
\sum_{i=1}^3\omega_i=n_--n_+ 
\end{equation}
is generically non-zero. It turns out that these 3-point functions vanish if $|n_+ - n_-|> 1$. In fact, one of the remarkable properties of the spectral flow operator is that the violation of the total winding number they induce is bounded: In a tree-level $n$-point scattering amplitude, the total winding number is bounded\footnote{In a genus-$g$ amplitude this bound is expected to be $|\sum_{i=1}^n\omega_i |\leq n-2+2g$.} by 
\begin{equation}
\Big \vert \sum_{i=1}^n\omega_i \Big \vert \leq n-2. \label{Bound}
\end{equation}
Amplitudes that do not obey this bound vanish identically. This bound, which was originally obtained by direct computation \cite{FZZ}, was explained in \cite{MO3} in terms of the symmetries of the WZW model. This is also consistent with the Ward identities derived in \cite{R}, and with the bounds obtained in the T-dual model \cite{FH}.

When considering (\ref{tormentita}), two main questions arise: 
\begin{itemize}
\item How to understand the presence of the spectral flow operator in a natural way from the CFT computation point of view\footnote{It was pointed out by Xi Yin that the presence of the spectral flow operator should be derivable from the bootstrap approach in the worldsheet CFT$_2$. Here, we will resort to the so-called $H_3^+$-Liouville correspondence, which follows from the identity between modular differential equations, and therefore this can be seen as an indirect realization of that idea.}. The way in which we will address this question involves the correspondence between WZW correlators such as (\ref{tormentita}) and correlators in LFT. This will lead us to make more precise the analysis of \cite{Nakayama}, where the spectral flow operators in relation to Liouville theory was first discussed. 
\item How to deal with multiple insertions of the spectral flow operator in a given string amplitude. In particular, how to deal with the accessory worldsheet marks where the spectral flow operators are inserted. Gauge invariance implies that the amplitude should not depend on those specific points $u^{\pm }_a$. The independence was proven for one or two spectral flow operators\footnote{Here, we will work in the so-called $m$-basis, in contrast to other works in which the spectral flow operator is written in the Mellin-transformed $x$-basis. In the $m$-basis, the independence from the inserting points $u_a^{\pm }$ becomes clearer.} \cite{MO3, Carmen}; here we want to prove this for an arbitrary number of them.
\end{itemize}

In answering these questions, we will proof the statements (a) and (b) above.

\section{$H_3^+$-Liouville correspondence redux}

To start our proof of (a) and (b), we have to work out the expression for $\tilde{C}^{3+n_++n_-}_{\text{AdS}_3}$. The first step to compute this correlator is to consider the so-called $H_3^+$-Liouville correspondence, which permits to write $SL(2,\mathbb{R})$ WZW $n$-point correlation functions\footnote{The $SL(2,\mathbb{R})$ WZW correlation functions, ${C}_{\text{AdS}_3}^{n}$, are usually defined by analytic extension from the correlators of the gauged $H^+_3=SL(2,\mathbb{C})/SU(2)$ WZW model. The discrete states appear as poles in the amplitudes, while the continuous series follow naturally from the normalizable states of the $H_3^+$ model.} in terms of $(2n-2+\omega )$-point LFT correlation functions, where $\omega $ is the total winding number in the WZW correlator. This correspondence follows from the relation existing between the solutions to the Knizhnik-Zamolodchikov (KZ) equation and the solutions to the Belavin-Polyakov-Zamolodchikov (BPZ) equation \cite{S}. This was revisited and generalized in \cite{RT, R}. In its general form, which includes the winding number, the $H_3^+$-Liouville correspondence formula reads\footnote{It corresponds to formula (3.29) in \cite{R} with $r=\omega $ and making $j_i\to -j_i$. This is also formula (3.36) of \cite{RT} making $m_i\to -m_i$, $j_i\to -j_i$, and setting $\omega_i=0$. A generalized version of the formula for arbitrary genus-$g$ and $\omega_i =0$ was given in \cite{HS}, where the $n$-point functions in the $SL(2,\mathbb{R})$ WZW model are in correspondence with the $(2n-2+2g)$-point functions of LFT.}
\begin{eqnarray}
&&C^{n}_{\text{AdS}_3}(z_1, ..., z_n)= \prod_{i=1}^n \frac{\Gamma(j_i-m_i)}{\Gamma(1-j_i+\bar{m}_i)} \ \sum_{\omega =0}^{n-2} \
\delta\Big(\sum_{i=1}^n\omega_i - \omega  \Big)\ \delta^{(2)}\Big(\sum_{i=1}^nm_i+k\omega /2\Big)  \nonumber \\
&&  \ \ \ \  \frac{2\pi^{3-2n}b\ (c_b)^{\omega }}{\Gamma(n-1-\omega )}\ \int_{\ }\prod_{a=1}^{n-2-\omega }d^2y_a \ \prod_{1\leq i<i'}^{n} \Big( (z_i-z_{i'})^{\beta_{ii'}-2\kappa_i \kappa_{i'}}
(\bar{z}_i-\bar{z}_{i'})^{\bar{\beta}_{ii'}-2\bar{\kappa }_i \bar{\kappa }_{i'}} \Big) \       \label{Ribault} \\
&&  \ \ \Big\langle  \prod_{i=1}^n e^{i\sqrt{2}(\kappa_i\chi(z_i)+\bar{\kappa}_i\bar{\chi}(\bar{z}_i))}\prod_{a=1}^{n-2-\omega } 
e^{i\sqrt{k/2}(\chi(y_a)+\bar{\chi}(\bar{y}_a))}
\Big\rangle_{\text{free}} \times  \Big\langle \prod_{i=1}^nV_{\alpha_i}(z_i) \prod_{a=1}^{n-2-\omega }V_{-\frac{1}{2b}}(y_a) \Big\rangle_{\text{LFT}} \nonumber
\end{eqnarray}
where $\kappa_i=m_i/\sqrt{k}-\sqrt{k}/2$, $\bar{\kappa}_i=\bar{m}_i/\sqrt{k}-\sqrt{k}/2$, $\beta_{ii'}=(1-\omega_i\omega_{i'}){k}/{2}-\omega_i{m}_{i'}-\omega_{i'}{m}_{i}-{m}_{i}-{m}_{i'}$, $\bar{\beta}_{ii'}=(1-\omega_i\omega_{i'}){k}/{2}-\omega_i\bar{m}_{i'}-\omega_{i'}\bar{m}_{i}-\bar{m}_{i}-\bar{m}_{i'}$, and where $\delta^{(2)}(z)=\delta(z)\delta(\bar{z})$. $c_b$ is a $b$-dependent factor that is unimportant here. We are assuming $\sum_{i=1}^{n}\omega_i \geq 0$, without loss of generality\footnote{To be more precise, the step $\omega = n-2$ of formula (\ref{Ribault}) was not proven in \cite{R} but it was presented as an educated conjecture, the reason being that in that case there are no degenerate fields in the LFT correlator an so its form can not be constrained by the BPZ equation. The case $\omega=n-2$ was later proven in \cite{violating} using free field techniques.}; an analogous formula exists for $\sum_{i=1}^{n}\omega_i \leq 0$. The integrals are defined on the whole $\mathbb{C}$-plane. The first expectation value on the third line of (\ref{Ribault}) is a ($2n-2+\omega $)-point correlation function of primary fields in the theory of a free\footnote{It can also be realized as an $n$-point correlation function of a $c<1$ matter CFT coupled to LFT, where the operators $e^{i\sqrt{k/2}\chi}\times V_{-1/(2b)}$ act as screening charges; this also explains the factor $c_b^{\omega }/\Gamma(n-1-\omega )$ as coming from combinatorics.} boson $\chi(z,\bar{z})=\chi(z)+\bar{\chi}(\bar{z})$. The second expectation value on the third line of (\ref{Ribault}) is an ($2n-2+\omega $)-point function of exponential primary operators in LFT; namely
\begin{equation}
\Big\langle \prod_{i=1}^n V_{\alpha _i}(z_i) \Big\rangle_{\text{LFT}} = \int \mathcal{D}\varphi \ e^{-S_{\text{L}}[\varphi ]} \prod_{i=1}^n e^{\sqrt{2}\alpha_i \varphi (z_i,\bar{z}_i)}\label{LFT1}
\end{equation}
where the Liouville action is
\begin{equation}
S_{\text{L}}[\varphi ]= \frac{1}{2\pi }\int d^2z \Big( \partial \varphi \bar{\partial } \varphi + \frac{Q}{4}R\varphi +\mu e^{\sqrt{2}b\varphi}\Big)  \ , \ \ \ Q=b+{1}/{b} \ . \label{LFT2}
\end{equation}
The relation between the quantum numbers in the WZW theory and in LFT is
\begin{equation}
\alpha_i=b({k}/{2}-j_i) \ , \ \ \ \ \  \ \ b^2={1}/({k-2})\ , 
\ \ \  \ \  \ \ \mu =b^2 , \label{Dic}
\end{equation}
with $i=1,2,..., n$. The constant $\mu $ is the Liouville cosmological constant, which here is fixed in terms of $b$. The constant parameter $b$ enters in the LFT central charge as $c=1+6Q^2$. For concreteness, let us consider the case of continuous representations $j_i=1/2-ip_i$ with $p_i\in \mathbb{R}$ for $i=1,2,3$, although exactly the same steps hold for arbitrary $SL(2,\mathbb{R})$ representations. Notice that the relation (\ref{Dic}) between the $SL(2,\mathbb{R})$ isospin $j_i$ and the LFT momenta $\alpha_i$ is such that states of the continuous representation $\mathcal{C}_{j\in 1/2+i\mathbb{R}}^{\lambda , \omega}$ get mapped into the normalizable states of LFT, namely those with $\alpha_i\in Q/2+i\mathbb{R}$.
\begin{figure}
\ \ \ \  \ \ \ \includegraphics[width=5.7in]{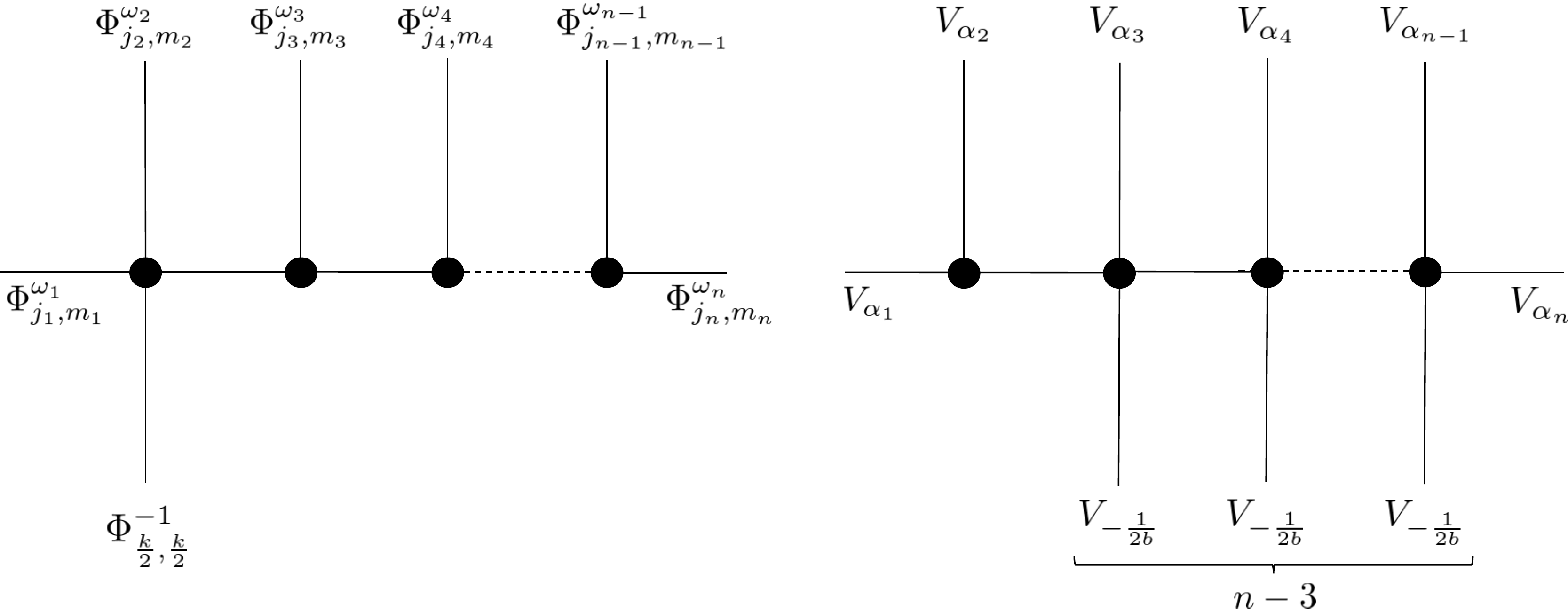}
\caption{Scheme of the $H_3^+$-Liouville correspondence. On the left, an $n$-point string scattering amplitude in AdS$_3$ in which the total winding number is $\omega\equiv\omega_1+\omega_2+...+\omega_n=1$; it is defined as an $(n+1)$-point correlation function in the $SL(2,\mathbb{R})_k$ WZW model involving one (as $\omega =1$) spectral flow operator. On the right, the corresponding $(2n-3)$-point correlation function in LFT involving $n-2-\omega $ degenerate fields $V_{-1/(2b)}$ integrated on the sphere.}
\label{Figure1}
\end{figure}

Out of the $2n-2-\omega $ operators in the LFT correlator of (\ref{Ribault}), $n-2-\omega $ of them have momentum $\alpha=-{1}/{(2b)}$; see Figure 1. This special value of the momentum corresponds to non-normalizable, degenerate states in LFT, meaning that $n-2-\omega $ states in the LFT correlator contain null descendants in the Verma modulo. The number of degenerate fields $V_{-1/(2b)}$ controls the violation of the winding number. This number goes from $0$ ($\omega =n-2$) to $n-2$ ($\omega =0$). In this formula, the bound $\omega \leq n-2$ comes from the Ward identities (2.31) of reference \cite{R}, whose solution is the product of $\omega +1$ $\delta $-functions in equation (3.26) therein\footnote{Using the notation of \cite{R}, if $\omega  \geq n-1$, one finds at least $n$ delta functions for $n$ variables $\mu_i$, which together impose $\mu_i = 0$, i.e. the $n$-point function with $\omega \geq n-1$vanishes almost everywhere in $\mu$-space, in agreement with inequality (\ref{Bound}) herein. The author thanks Sylvain Ribault for clarifying this point.}.

We can use the $H_3^+$-Liouville correspondence formula (\ref{Ribault}) to write the correlator (\ref{tormentita}) as
\begin{eqnarray}
\tilde{C}^{\ 3,n_+ , n_-}_{\text{AdS}_3}  &=& \lim _{\epsilon \to 0}\frac{(\Gamma(\epsilon ))^{n_-}}{Z_{+, \epsilon}^{n_+}Z_{-, \epsilon}^{n_-}}
\frac{2(\Gamma(k))^{n_+}\pi^{-3-2n_+-2n_-}}{\sqrt{k-2}(\Gamma(1-k))^{n_+}\Gamma(2+n_++n_-)}\prod_{a=1}^{3}
\frac{\Gamma(\frac 12-ip_a-m_a)}{\Gamma(\frac 12+ip_a+m_a)}\nonumber \\
&&  \ \ \ \ \ \ \delta^{(2)}\Big(\sum_{a=1}^3m_a-\frac k2(n_+-n_-)\Big) 
 \prod_{a=1}^{n_+} \Big( |u^+_a|^{2k-4m_1} |1-u^+_a|^{2k-4m_2} \Big)  \label{tornadito} \\
&&   \  \prod_{1\leq a<a'}^{n_+} |u^+_a-u^+_{a'}|^{4k} \int_{\ } \prod_{l=1}^{n_++n_-+1} d^{2}y_l \prod_{1\leq l <l'}^{n_++n_-+1} |y_l-y_{l'}|^{k}
\prod_{a=1}^{n_+}\prod_{l=1}^{n_++n_-+1} |u^+_a-y_l|^{-2k} \nonumber \\
&&  \prod_{l=1}^{n_++n_-+1} \Big( |y_l|^{2m_1-k} |1-y_l|^{2m_2-k} \Big) \ \Big\langle V_{\alpha_1}(0) V_{\alpha_2}(1) V_{\alpha_3}(\infty ) \prod_{l=1}^{n_++n_-+1}V_{-\frac{1}{2b}}(y_l) \Big\rangle_{\text{LFT}} \nonumber
\end{eqnarray} 
where $\alpha_i=b(-j_i+b^{-2}/2+1)=(k/2-1/2+ip_i)/\sqrt{k-2}$, $i=1,2,3$, with $j_i=1/2-ip_i$, $p_i\in \mathbb{R}$, and $b=1/\sqrt{k-2}$. This implies that for the remaining operators we have $\alpha_{a>3}=0$ with $a=4,5,... ,n_-+n_+$. 

We already see some magic working here: All the dependence on the variables $u^-_A$ have disappeared from the expression above! We will see next that, by rewriting this expression in a smart way, also the remaining variables $u^+_a$ are seen to be superfluous. This is not evident from (\ref{tornadito}) but it will become clear after some algebraic manipulation.

In the expression above, the subscript $\epsilon $ in the normalization factors $Z_{\pm , \epsilon}$ appears because we want to use the freedom in the normalization of the spectral flow operators to eventually cancel the divergent factor $(\Gamma(\epsilon \to 0))^{n_-}$ together with other divergences that might eventually appear through the calculation.   

From (\ref{tormentita}), we have $\sum_{i=1}^3\omega_i+n_+-n_-=0$. We will consider the case $n_+-n_-=-1$, which corresponds to $\sum_{i=1}^3\omega_i=1$. It involves the contribution of the $n_++n_-=2n_--1$ spectral flow operators, with $n_-$ being arbitrary. In other words, it corresponds to $\sum_{i=1}^3m_i=\sum_{i=1}^3\bar{m}_i=(n_+-n_-)k/2=-k/2$. We also assume $m_i = \bar{m}_i$ for simplicity.

Our strategy will be the following: $1)$ We will start with the expression (\ref{tornadito}) for $\tilde{C}^{3+n_-+n_+}_{\text{AdS}_3}$, in which we used the step $\omega =0$ of the formula (\ref{Ribault}) to express $\tilde{C}^{3+n_-+n_+}_{\text{AdS}_3}$ in terms of a LFT correlator. $2)$ We will work out the expression as much as we can using a series of integral identities and duality relations. $3)$ We will resort again to formula (\ref{Ribault}) and use the step $\omega =1$ of the sum to verify that, when $n_-=n_++1$, the original quantity $\tilde{C}^{3+n_-+n_+}_{\text{AdS}_3}$ actually coincides with the $SL(2,\mathbb{R})$ WZW 3-point amplitude that violates the winding number in $+1$. In other words, we will see that the step $\omega =0$ of the sum appearing in formula (\ref{Ribault}), when $n_--n_+=1$ spectral flow operators are considered, coincides with the step $\omega =1$ of the same formula (\ref{Ribault}) when no spectral flow operators are present. That is to say, we will prove the consistency between the FZZ prescription and the $H_3^+$-Liouville correspondence; see (\ref{Borron}) below.

\section{Integral representation and duality}

In order to get a closed expression for $\tilde{C}^{3, n_+, n_-}_{\text{AdS}_3}$, we still need to solve the LFT correlators on the right hand side of (\ref{tornadito}). To do so, we could try with the standard Coulomb gas realization \cite{GL}, namely
\begin{equation}
\Big\langle \prod_{i=1}^{\tilde{n}}V_{\alpha_i}(z_i)  \Big\rangle_{\text{LFT}} = \mu^{\tilde{s}} \Gamma(-{\tilde{s}})\prod_{1\leq i<i'}^{\tilde{n}} |z_i-z_{i'}|^{-4\alpha_i\alpha_{i'}} \ I_{\tilde{s}}(\alpha_1 , ..., \alpha_{\tilde{n}}; z_1 , ..., z_{\tilde{n}}; b) \label{integral}
\end{equation}
with 
\begin{equation}
I_{\tilde{s}}(\alpha_1 , ..., \alpha_{\tilde{n}}; z_1 , ..., z_{\tilde{n}}; b) = \int_{\ } \prod_{r=1}^{\tilde{s}} d^2w_r 
\prod_{i=1}^{\tilde{n}} \prod_{r=1}^{\tilde{s}} |z_i - w_r|^{-4b\alpha_i} \prod_{1\leq r<r'}^{\tilde{s}}|w_r - {w}_{r'}|^{-4b^2}\label{Nueva}
\end{equation}
where ${\tilde{s}}=1+b^{-2}-b^{-1}\sum_{i=1}^{\tilde{n}}\alpha_i$ and where the integrals are on the whole $\mathbb{C}$-plane. --The factor $\mu^{\tilde{s}} \Gamma(-\tilde{s})$ in (\ref{integral}), together with the $\tilde{s}$ insertions at $w_1,w_2,... , w_{\tilde{s}}$ come from the integration over the zero mode of the Liouville field $\langle \varphi \rangle = \varphi -\delta \varphi$, and $\mu $ is the Liouville cosmological constant.-- However, although realization (\ref{integral}) turns out to be useful in many contexts, it would not be of help here. Instead, it is convenient to resort to its dual realization, which amounts to use the self-duality that LFT exhibits under $b \leftrightarrow  b^{-1}$. This permits to alternatively express the LFT correlators as follows
\begin{equation}
\Big\langle \prod_{i=1}^{\tilde{n}}V_{\alpha_i}(z_i)  \Big\rangle_{\text{LFT}} = \tilde{\mu}^s \Gamma(-s)\prod_{1\leq i<i'}^{\tilde{n}} |z_i-z_{i'}|^{-4\alpha_i\alpha_{i'}} 
\ I_{s}(\alpha_1 , ..., \alpha_{\tilde{n}}; z_1 , ..., z_{\tilde{n}}; 1/b) \label{integrala}
\end{equation}
where $s=b^2\tilde{s}$ and $\tilde{\mu} = \mu^{-b^{-2}}\pi^{b^{-2}-1}\Gamma(1-b^{-2})(\Gamma(b^2))^{b^{-2}}(\Gamma(1-b^2))^{-b^{-2}}(\Gamma(b^{-2}))^{-1}$. Here, because of (\ref{Dic}), we have a particular value of $\mu $ and therefore of $\tilde{\mu }$. Nevertheless, we prefer to write the expressions for generic $\tilde{\mu }$ as it permits to keep track of the Knizhnik-Polyakov-Zamolodchikov scaling of the correlators and to have control over the string coupling constant on AdS$_3$ \cite{Notes}. 

The ones appearing on the right hand side of (\ref{integrala}) are $s$ Selberg-type integrals over $\mathbb{CP}^1\backslash \{ z_1, ... ,z_n\}$. The measure is $ d^2w_r = (i/2)dw_r d\bar{w}_r$ with, say, $w_r = x_r + i y_r$, $\bar{w}_r = {x}_r - i {y}_r$. To solve (\ref{integrala}) it is convenient to Wick-rotate $x_r\to ix_r$ and introduce a deformation parameter in $|w_r |^2=-x_r ^2+y_r ^2+i\varepsilon $ to avoid the poles at $x_r =\pm y_r $. The way to proceed is first to define coordinates $x^{\pm }_r = \pm  x_r + y_r $ and then integrate over $x^-_r$ while keeping $x^+_r$ fixed\footnote{The author thanks Mat\'{\i}as Leoni for discussions about this point.}.

Of course, the integral formula (\ref{integral}) only makes sense for kinematical configurations such that $\tilde{s}\in \mathbb{Z}_{\geq 0}$. This is because $s$ in (\ref{Nueva}) is the number of integrals to be performed! Nevertheless, one can make sense out of (\ref{integrala}) in more general cases: To do so, one first assumes $s\in \mathbb{Z}_{\geq 0}$, then solves the integrals in terms of elementary functions, and finally analytically continues the expressions to $s\in \mathbb{C}$. Here, throughout the formulae, we will assume this kind of analytic extension.

Using the integral realization (\ref{integrala}), we get the following expression for $\tilde{C}_{\text{AdS}_3}^{\ 3, n_--1,n_-}$ 
\begin{eqnarray}
&&\tilde{C}^{\ 3,n_--1 , n_-}_{\text{AdS}_3}=
\lim_{\epsilon \to 0}\frac{(\Gamma(\epsilon ))^{n_-}}{Z_{+,\epsilon}^{n_--1}Z_{-,\epsilon}^{n_-}}
\frac{2\Gamma(-s)(\Gamma(k))^{n_- -1}\tilde{\mu}^{s} \pi^{-4n_- -1}}{\sqrt{k-2}(\Gamma(1-k))^{n_- -1}\Gamma(2n_- + 1)}\prod_{a=1}^{3}
\frac{\Gamma(\frac 12-ip_a-m_a)}{\Gamma(\frac 12+ip_a+m_a)}\nonumber \\
&&
\ \ \ \ \  \delta^{(2)}\Big(\sum_{a=1}^3m_a+\frac k2\Big)
\int_{\ } \prod_{l=1}^{2n_- } d^2{y}_{l} \int_{\ } \prod_{r=1}^{s} d^2{w}_{r}
\prod_{1\leq l <l'}^{2n_- } |y_l-y_{l'}|^{2} \prod_{1\leq r < r'}^{s} |w_r-w_{r'}|^{8-4k}
\nonumber \\
&&
 \prod_{l=1}^{2n_- } \prod_{r=1}^{s} |y_l-w_r|^{2k-4}
\prod_{1\leq a<a'}^{n_- -1} |u^+_a-u^+_{a'}|^{4k} \prod_{a=1}^{n_- -1}\prod_{l=1}^{2n_- } |u^+_a-y_l|^{-2k}
\prod_{a=1}^{n_- -1} \Big( |u^+_a|^{2k-4m_1} |1-u^+_a|^{2k-4m_2} \Big) \nonumber \\
&& \ \ \ \ \prod_{l=1}^{2n_-} \Big( |y_l|^{2m_1-1+2ip_1} |1-y_l|^{2m_2-1+2ip_2} \Big) \prod_{r=1}^{s} \Big( |w_r|^{2-4ip_1-2k} |1-w_r|^{2-4ip_2-2k} \Big) \label{cumersome}
\end{eqnarray} 
with $s(k-2)=-\sum_{i=1}^3\alpha_i/b+(1+n_+ + n_-)/(2b^2)+1+1/b^2=-i\sum_{i=1}^3p_i-1/2+(k/2-1)(2n_- -1)$. Notice the factor $(\Gamma(1-k))^{1-n_-}$ in the expression above, which for $k\in \mathbb{Z}_{>2}$ represents a factor zero for $n_->1$. This factor can be eventually absorbed in the normalization $(Z_{+,\epsilon})^{1-n_-}$, together with other factors; see below. The factor $\Gamma(-s)$, which is divergent for $s\in \mathbb{Z}_{\geq 0}$, is the well-known factor that accompanies the resonant correlators in the Coulomb gas realization; this leads to the residues of the observables with the appropriate factor once one uses the relation $\Gamma(-s)\sim \Gamma(0) (-1)^s/s!$ and isolates the single pole.

Expression (\ref{cumersome}) is cumbersome and somehow obscure: In particular, it does not make explicit the fact that $\tilde{C}_{\text{AdS}_3}^{\ 3, n_--1,n_-}$ does not depend on the variables $u^+_a$. Therefore, it is still necessary to simplify it further. To do so, we can resort to the integral relation\footnote{As a consistency check, we can use integral formula (\ref{La19}) together with the $H_3^+$-Liouville correspondence formula (\ref{Ribault}) to write the resonant WZW $n$-point correlators with momenta satisfying $\sum_{i=1}^n j_i=n-1$ and $ \sum_{i=1}^n \omega_i = 0$ and show they take the form $C_{\text{AdS}_3}^n \propto \prod_{1\leq i <j}^n |z_i - z_j|^{-4t_{ij}}$ with $t_{ij}=(1-j_i)(1-j_j)/(k-2)+(m_i+k\omega_i /2)(m_i+k\omega_i /2)-m_i m_j $. This expression can be derived independently, for example using the Wakimoto free field representation.}
 \cite{Basilea, FL}
\begin{eqnarray}
&&\int_{\ }\prod_{i=1}^{n}d^2y_i \prod_{1\leq 1 <i'}^{n} |y_i - y_{i'}|^2 \prod_{i=1}^{n}\prod_{i=1}^{n+m+1} |y_i - x_j|^{2L_{j}} = \pi^{n-m} \frac{\Gamma(n+1)}{\Gamma(m+1)}\frac{\Gamma(-n-\sum_{j=1}^{n+m+1}L_j)}{\Gamma(1+n+\sum_{j=1}^{n+m+1}L_j)}   \nonumber \\
&&\ \ \prod_{j=1}^{n+m+1} \frac{\Gamma(1+L_j)}{\Gamma(-L_j)} \prod_{1\leq j <j'}^{n+m+1}|x_j - x_{j'}|^{2+2L_j+2L_{j'}} \int_{\ }\prod_{l=1}^{m}d^2y_l \prod_{1\leq l<l'}^{m} |y_{l}-y_{l'}|^2  \prod_{l=1}^{m} \prod_{j=1}^{n+m+1} |y_l - x_j|^{-2-2L_j} . \nonumber \\ \label{La19}
\end{eqnarray}
We can use this formula to change the number of integrals over the variables $y_l$ from $n_++n_-+1=2n_-$ to $s-n_-$. This amounts to consider in (\ref{La19}) the case $n=2n_-$, $m=s-n_-$. This yields
\begin{eqnarray}
&&\tilde{C}^{\ 3,n_- -1 , n_-}_{\text{AdS}_3}=
-\lim_{\epsilon \to 0} \frac{(\Gamma(\epsilon ))^{n_- }}{Z_{+,\epsilon }^{n_--1}Z_{-,\epsilon }^{n_-}}
\frac{2\Gamma(-s)\pi^{-1-n_- -s}}{\sqrt{k-2}\Gamma(s-n_-+1)}
\Big(\tilde{\mu}\frac{\Gamma(k-1)}{\Gamma(2-k)}\Big)^{s} 
\nonumber \\
&&\ \ \ \  \delta^{(2)}\Big(\sum_{a=1}^3m_a+\frac k2\Big) \int_{\ } \prod_{l=1}^{s-n_-} d^2{y}_{l} \int_{\ } \prod_{r=1}^{s} d^2{w}_{r}
\prod_{1\leq l <l'}^{s-n_-} |y_l-y_{l'}|^{2} 
\prod_{1\leq r < r'}^{s} |w_r-w_{r'}|^{2} 
\nonumber \\
&& \ \ \ \ 
\prod_{l=1}^{s-n_-} \prod_{r=1}^{s} |y_l-w_r|^{2-2k}
\prod_{r=1}^{s}\prod_{a=1}^{n_- -1}|w_r-u^+_a|^{-2}\prod_{1\leq a<a'}^{n_- -1} |u^+_a-u^+_{a'}|^{2} \prod_{a=1}^{n_- -1}\prod_{l=1}^{s-n_-} |u^+_a-y_l|^{2k-2}
\nonumber \\
&& \ \ \ \ 
\prod_{a=1}^{n_- -1} \Big( |u^+_a|^{1+2ip_1-2m_1} |1-u^+_a|^{1+2ip_2-2m_2} \Big) 
 \prod_{l=1}^{s-n_-} \Big( |y_l|^{-1-2ip_1-2m_1} |1-y_l|^{-1-2ip_2-2m_2} \Big) 
\nonumber \\
&& \ \ \ \ 
\prod_{r=1}^{s} \Big( |w_r|^{-1+2m_1-2ip_1} |1-w_r|^{-1+2m_2-2ip_2} \Big).
\end{eqnarray}
This expression, although still complicated, exhibits some promising simplifications with respect to (\ref{cumersome}). In particular, we notice that the exponent $2$ in the factors $|w_r-w_{r'}|^{2}$ permits to use formula (\ref{La19}) again to integrate out the variables $w_r$. Considering in (\ref{La19}) the case $n=s$, $m=0$, the expression for $\tilde{C}^{\ 3,n_- -1 , n_-}_{\text{AdS}_3}$ reduces to
\begin{eqnarray}
&&\tilde{C}^{\ 3,n_- -1 , n_-}_{\text{AdS}_3}=\lim_{\epsilon \to 0}\frac{(\Gamma(\epsilon ))^{2n_--1}}{ Z_{+,\epsilon }^{n_- -1}Z_{- ,\epsilon }^{n_-}} \Big( - \frac{\tilde{\mu}}{\pi}\frac{\Gamma(k-1)}{\Gamma(2-k)}\Big)^{n_-}\prod_{a=1}^{3}\frac{\Gamma(\frac 12 -ip_a+m_a)}{\Gamma(\frac 12 +ip_a-m_a)} \delta^{(2)}\Big(\sum_{a=1}^3m_a+\frac k2\Big)
\nonumber \\
&&
\ \ \frac{2bc_b}{\pi^3}\tilde{\mu}^{s-n_-}\Gamma(n_--s)\int_{\ } \prod_{l=1}^{s-n_-} d^2{y}_{l}  
\prod_{1\leq l <l'}^{s-n_-} |y_l-y_{l'}|^{8-4k} \prod_{l=1}^{s-n_-} 
 \Big( |y_l|^{4ip_1+2-2k} |1-y_l|^{4ip_2+2-2k} \Big).\nonumber
\end{eqnarray}
where we have used that $\Gamma(-s)\Gamma(s+1) = (-1)^{n_-} \Gamma(s-n_-+1)\Gamma(n_--s)$. $c_b$ stands for an irrelevant factor, which depends on $b$ but does not depend neither on the momenta $p_i$ nor on $m_i$. 

Remarkably, as we anticipated, the variables $u^+_a$ have totally disappeared from the expression! Moreover, using (\ref{integrala}) we can recognize the second line of the last expression as the integral representation of a LFT 3-point function. This yields, 
\begin{eqnarray}
\tilde{C}^{\ 3,n_+ , n_-}_{\text{AdS}_3}= \frac{2bc_b}{\pi^3} \prod_{a=1}^{3}\frac{\Gamma(\frac 12 -ip_a+m_a)}{\Gamma(\frac 12 +ip_a-m_a)} \ \delta^{(2)}\Big(\sum_{a=1}^3m_a+\frac k2\Big)
  \ \Big\langle \prod_{i=1}^3 V_{\alpha_i}(z_i)\Big\rangle_{\text{LFT}} ,\label{UIO}
\end{eqnarray}
where $\alpha_i=b(k/2-1/2-ip_i)$. We have absorbed in the normalization some factors; more precisely, we fixed $Z_{+,\epsilon}=\Gamma (\epsilon )$ and $Z_{-,\epsilon }=Z_{+,\epsilon } \tilde{\mu}\Gamma(b^{-2})/(b^4\pi \Gamma(1-b^{-2}))  $ --this can actually be seen as a consistency check, since it was not a priori obvious that both the dependence on the arbitrary number $n_-$ and the divergences could have been absorbed in the normalization factors--.  

We can express (\ref{UIO}) in terms of the $\Upsilon$-functions introduced in \cite{ZZ} by simply evaluating the Dorn-Otto-Zamolodchikov-Zamolodchikov formula for the LFT 3-point function. Besides, resorting to the functional identity\footnote{This function is defined in terms of the Barnes double Gamma function, $\Gamma_2(a| b,c)$, as follows: $G(j) =  \Gamma_2 (-j | 1, k-2)\Gamma_2 (k-1+j | 1, k-2) (k-2)^{j(k-1-j)/(2k-4)}$. See (2.13)-(2.16) in \cite{MO3} and references therein.} $G(x)=b^{-b^2x^2-(b^2+1)x}\Upsilon_{b}^{-1}(-bx)$, we can easily show that expression (\ref{UIO}) reproduces the results of \cite{GN3, MO3, Giribet, violating} for the winding-violating 3-point amplitude in AdS$_3$. 

Alternatively, we can revert the argument and use again formula (\ref{Ribault}) to verify that the right hand side of (\ref{UIO}) is nothing but the winding-violating WZW 3-point function itself. Then, we conclude
\begin{eqnarray}
\Big\langle \prod_{i=1}^{3} \Phi_{\frac 12 -p_i,m_i,\bar{m}_i}^{\omega_i} \prod_{a=1}^{n_--1} \Phi_{\frac k2,-\frac k2,-\frac k2}^{+1} \prod_{A=1}^{n_-} \Phi_{\frac k2,+\frac k2,+\frac k2}^{-1} \Big\rangle_{sl(2)}^{0}={Z_+^{n_--1}Z_-^{n_-}}
\Big\langle \prod_{i=1}^{3} \Phi_{\frac 12 +p_i,-m_i,-\bar{m}_i}^{\omega_i}  \Big\rangle_{sl(2)}^{1}
\label{Borron}
\end{eqnarray} 
This is precisely the identity we wanted to prove: The $(2+2n_-)$-point function $\tilde{C}_{\text{AdS}_3}^{3,n_--1,n_-}$ actually gives the 3-point function $C^3_{\text{AdS}_3}$ with total winding number $\sum_{i=1}^3 \omega _i =1$. 

\section{The $n$-point function}

Expression (\ref{UIO})-({\ref{Borron}}) manifestly show the independence from the inserting points of the auxiliary spectral flow operators, i.e. ${\partial}_{u^{\pm }_a}\tilde{C}^{3, n_--1 ,  n_-}_{\text{AdS}_3}=0$. This is far from evident if one starts, for example, with representation (\ref{cumersome}), which seemed to depend at least on the $u^+_a$. Furthermore, we argued that, after having chosen the adequate normalization, the result do not depend on $n_-$ and $n_+$ separately, but it depends on the difference $n_--n_+$. In other words, the vertices $\Phi_{k/2,k/2,k/2}^{-1}$ and $\Phi_{k/2,-k/2,-k/2}^{1}$ neutralize each other. 

Going back to the independence from $u^{\pm}_a$, we notice that a similar phenomenon happens with $n$-point functions: Consider for example the $n$-point function 
\begin{eqnarray}
\tilde{C}^{\ n, 0 ,  n_-}_{\text{AdS}_3}(z_1, ... , z_n)=\Big\langle \prod_{i=1}^{n} \Phi_{j_i,m_i,\bar{m}_i}^{\omega_i}(z_i)  \prod_{a=1}^{n_-} \Phi_{\frac k2,+\frac k2,+\frac k2}^{-1}(u^-_a) \Big\rangle_{sl(2)}^{0} .
\end{eqnarray} 
Using (\ref{Ribault}) and (\ref{integrala}), this yields
\begin{eqnarray}
&&\tilde{C}^{\ n, 0 ,  n_-}_{\text{AdS}_3}(z_1, ... , z_n)=\lim _{\epsilon \to 0}
\frac{(\Gamma(\epsilon ))^{n_-}}{Z_{-,\epsilon }^{n_-}}
\frac{2\Gamma(-s)\tilde{\mu}^{s} \pi^{-2n_- -3}}{\sqrt{k-2}\Gamma(n-1+n_-)} \prod_{1\leq i<j}^{n} |z_i - z_j |^{2\hat{\beta}_{ij}} \prod_{a=1}^{n}
\frac{\Gamma(\frac 12-ip_a-m_a)}{\Gamma(\frac 12+ip_a+m_a)}\nonumber \\
&& \ \ \ \ \ 
\delta^{(2)}\Big(\sum_{a=1}^3m_a+\frac{k}{2}n_-\Big)
\int_{\ } \prod_{l=1}^{n_- +n-2} d^2{y}_{l} \int_{\ } \prod_{r=1}^{s} d^2{w}_{r}
\prod_{1\leq l <l'}^{n_- +n-2} |y_l-y_{l'}|^{2} \prod_{1\leq r < r'}^{s} |w_r-w_{r'}|^{8-4k}
\nonumber \\
&& \ \ \ \ \ \prod_{l=1}^{n_- +n-2} \prod_{r=1}^{s} |y_l-w_r|^{2k-4}   \prod_{a=1}^{n}  \prod_{l=1}^{n_- +n-2}  |z_a - y_l|^{2m_a-1+2ip_a} \ \prod_{b=1}^{n} \prod_{r=1}^{s}  |z_b - w_r|^{2-4ip_b-2k}  \nonumber
\end{eqnarray} 
with $s(k-2)=-i\sum_{i=1}^np_i+1-n/2+(k/2-1)n_-$ and $\hat{\beta}_{ij}= \beta_{ij}-2\alpha_i\alpha_j$. This expression is actually independent from $u_a^-$, although it does depend on $n_-$ yielding $\sum_{i=1}^n\omega_i = n_-$. The independence from the variables $u^{-}_a$ is here observed at the level of $n$-point correlation functions, and not only at the level of correlation numbers or amplitudes.

\subsection*{Acknowledgments}

The author is indebted to Matthew Kleban and Massimo Porrati for many discussions and ideas. He is also grateful to Christopher Hull, Mat\'{\i}as Leoni, Yu Nakayama, and Eliezer Rabinovici for enjoyable collaborations in related subjects. He has also been benefited by discussions with Sylvain Ribault and Xi Yin. Finally, he thanks the hospitality of the following institutions, where different parts of this work were done: the Abdus Salam International Centre for Theoretical Physics, the Center for Cosmology and Particle Physics of New York University, the Simons Center for Geometry and Physics at Stony Brook University.

  \end{document}